\documentclass[12pt,preprint]{aastex}

\def\pf{p_f}

\def\rf{r_f}

\def\Gej{\Gamma_{\rm ej}}

\def\Lej{L_{\rm ej}}

\def\tobs{t_{\rm obs}}

\def\bargp{\bar \gamma_p}
\def\gm{\gamma_m}
\def\gc{\gamma_c}
\def\ginj{\gamma_{\rm inj}}

\def\nuobs{\nu_{\rm obs}}

\def\be{\begin{equation}}
\def\ee{\end{equation}}
\def\beq{\begin{eqnarray}}
\def\eeq{\end{eqnarray}}

\begin{document}

\title{On the Non-existence of a Sharp Cooling Break in GRB Afterglow Spectra} 

\author{Z. Lucas Uhm\altaffilmark{1,2}, Bing Zhang\altaffilmark{1,3,2}}

\altaffiltext{1}{Kavli Institute for Astronomy and Astrophysics, Peking University, Beijing 100871, China}

\altaffiltext{2}{Department of Physics and Astronomy, University of Nevada, Las Vegas, NV 89154, USA; uhm@physics.unlv.edu, zhang@physics.unlv.edu}

\altaffiltext{3}{Department of Astronomy, Peking University, Beijing 100871, China}

\begin{abstract}
Although the widely-used analytical afterglow model of gamma-ray bursts (GRBs) 
predicts a sharp cooling break $\nu_c$ in its afterglow spectrum, 
the GRB observations so far rarely show clear evidence for a cooling break 
in their spectra or its corresponding temporal break in their light curves. 
Employing a Lagrangian description of the blast wave, we conduct a sophisticated 
calculation of the afterglow emission. We precisely follow the cooling history of 
non-thermal electrons accelerated into each Lagrangian shell. We show that a detailed 
calculation of afterglow spectra does not in fact give rise to a sharp cooling break 
at $\nu_c$. Instead, it displays a very mild and smooth transition, which occurs gradually 
over a few orders of magnitude in energy or frequency. 
The main source of this slow transition is that different mini-shells have
different evolution histories of the comoving magnetic field strength $B$, so that 
deriving the current value of $\nu_c$ of each mini-shell requires an integration of 
its cooling rate over the time elapsed since its creation. 
We present the time evolution of optical and X-ray 
spectral indices to demonstrate the slow transition of spectral regimes, and discuss 
the implications of our result in interpreting GRB afterglow data.
\end{abstract}


%
%

\section{Introduction} \label{section:introduction}

The broad-band afterglow emission of gamma-ray bursts (GRBs) \citep{costa97,vanpara97},
has been interpreted as synchrotron radiation from a relativistic blast wave, which 
sweeps up a surrounding ambient medium \citep{meszarosrees97,sari98}. 
This widely-used analytical model of GRB afterglows \citep{sari98}
predicts a sharp cooling break at $\nu_c$ in their spectra, 
determined by synchrotron cooling rate of non-thermal electrons during
the dynamical time scale of the relativistic blast wave.

GRB afterglow observations so far rarely show clear evidence 
for a cooling break in their spectra or its corresponding temporal break 
in light curves with a spectral index change across the break. The Swift
satellite \citep{gehrels04} has accumulated 9 years of afterglow data.
In the X-ray band where $\nu_c$ likely shows up, usually one or two steepening
breaks are observed in the canonical X-ray light curves \citep{zhang06,nousek06}.
However, the change of decay slope is inconsistent with $\nu_c$ crossing,
and more importantly, there is essentially no spectral index change across 
the temporal breaks \citep[e.g.][]{liang07b,liang08}. Inferences of $\nu_c$
were only occasionally drawn in rare bursts \citep[e.g.][]{filgas11}.
The missing $\nu_c$ in individual GRBs
may be understood as that the shock parameters are such that $\nu_c$ lies in
between optical and X-ray bands, so that there is no
$\nu_c$ crossing during the observational time \citep[e.g.][]{curran10,oates11}.
However, the lack of $\nu_c$ in almost entire Swift afterglow data set 
suggests that this must be an effect intrinsic to GRB physics.

The simple analytical model \citep{sari98} assumes that
the entire postshock material forms a single zone with the same energy density and 
magnetic field. The single shocked zone is endowed with a broken power-law electron 
energy distribution with a break at a ``cooling'' Lorentz factor $\gamma_c$. 
\cite{granotsari02} relaxed this assumption and 
introduced other factors such as the Blandford-McKee (BM) solution \citep{blandford76},
the curvature effect, and adiabatic cooling to describe the shocked electrons, and showed that 
the shape of spectral breaks is much smoother than predicted 
by the analytical model (see also \cite{vaneerten09,leventis12}). 
However, since their model introduces many factors,
one could not identify the main source of smoothing.

In this paper, we perform a detailed study on the formation of GRB afterglow 
spectra, by adopting a Lagrangian description of the shocked region 
\citep[][hereafter U12]{uhm12}. We obtain a very smooth cooling break $\nu_c$ in the
afterglow spectrum. We identify its main physical origin, and show that
a smooth $\nu_c$ is ubiquitous and intrinsic to a wide range of 
astrophysical phenomena involving synchrotron cooling.

%
%

\section{Smooth cooling break and its physical origin} \label{section:cooling}

Following \cite{uhm11} and U12 (Section 2), 
we make use of a semi-analytic 
formulation of a relativistic blast wave in order to find its dynamics 
accurately from an initial coasting phase through a deceleration stage. 
Then, as described in U12 (Section 3), we adopt a Lagrangian description 
of the blast wave to conduct a sophisticated calculation of its afterglow 
emission. The blast is viewed as being made of many different Lagrangian 
shells $\{\delta m^i\}$. Here the index $i$ is used to denote each Lagrangian 
shell. We keep track of an adiabatic evolution of each shell $\delta m^i$ 
to find a time evolution of the magnetic field $B^i$ of the shell (U12, Section 3.1). 
The cooling history of an electron spectrum is also followed individually 
for every shell $\delta m^i$ (U12, Section 3.2).

Radiative and adiabatic cooling of an electron with the Lorentz factor $\gamma_e$ 
is described by the first and second term below, respectively, (U12, Equation (17)) 
\be
\label{eq:gamma_e}
\frac{d}{d t^{\prime}} \left(\frac{1}{\gamma_e}\right) = 
\frac{\sigma_T}{6\pi m_e c}\, B^{i\,2} - 
\frac{1}{4} \left(\frac{1}{\gamma_e}\right) \frac{d \ln p^i}{d t^{\prime}} .
\ee
Here $t^{\prime}$ is the time measured in the co-moving fluid frame, 
$\sigma_T$ is the Thomson cross section, $m_e$ is the electron mass, 
$c$ is the speed of light, and $B^i$ and $p^i$ are respectively the magnetic field 
and the pressure of the shell $\delta m^i$ where the electron sits in. 
The Compton parameter $Y$ that describes a contribution of inverse Compton 
scattering to the cooling rate of electrons is omitted here. We assume
that electrons are accelerated into a power-law distribution of slope $p$ 
above an injection Lorentz factor $\ginj$ in a fresh shell $\delta m^i$ 
at a shock front (forward shock (FS) or reverse shock (RS)). 
As the blast wave propagates, the electron spectrum in the shell $\delta m^i$ evolves 
in time. All the electrons inside the spectrum cool down, and their cooling 
history is governed by the differential Equation (\ref{eq:gamma_e}). 
Thus, we use Equation (\ref{eq:gamma_e}) to find a time evolution of 
a minimum Lorentz factor $\gm^i$ and a cooling Lorentz factor $\gc^i$ 
for the shell $\delta m^i$. For simplicity, we further assume that 
the shell $\delta m^i$ maintains a power-law electron spectrum 
of slope $p$, bounded by two Lorentz factors $\gm^i$ and $\gc^i$, 
as time goes.

More explicitly, when the shell $\delta m^i$ is created at a shock front (FS or RS) 
at the co-moving time $t_i^{\prime}$, 
the electron spectrum injected into the shell $\delta m^i$ has the initial 
Lorentz factors 
\be
\label{eq:initial}
\gm^i (t_i^{\prime}) = \ginj (t_i^{\prime}),
\qquad
\gc^i (t_i^{\prime}) = +\infty,
\ee
where
\be
\label{eq:injection}
\ginj (t^{\prime}) = 
1+\frac{p-2}{p-1}\, \frac{m_p}{m_e}\, \epsilon_e\, \left[\bargp (t^{\prime})-1\right].
\ee
Here, $m_p$ is the proton mass, 
and $\bargp (t^{\prime})$ is the mean Lorentz factor of protons in the fresh shell 
created at the shock front at co-moving time $t^{\prime}$. 
Then, the Lorentz factors $\gm^i (t_j^{\prime})$ and $\gc^i (t_j^{\prime})$ of 
the shell $\delta m^i$ at a later time $t_j^{\prime}$ ($>t_i^{\prime}$) can be 
found by integrating Equation (\ref{eq:gamma_e}) from $t_i^{\prime}$ to $t_j^{\prime}$.\footnote{
With initial values given by Equation (\ref{eq:initial}), 
$\gm^i (t_j^{\prime}) < \gc^i (t_j^{\prime})$ is always expected for the given shell $\delta m^i$. 
However, $\gc^i (t_j^{\prime}) < \gm^j (t_j^{\prime})$ can also happen for fast cooling, 
which corresponds to the fast cooling regime ($\gc < \gm$) in the context of \cite{sari98}.}

We take the following simple fiducial example to calculate the afterglow spectra.
(1) A constant density $\rho_1(r)/m_p=n_1(r)=1~\mbox{cm}^{-3}$ is assumed for the ambient medium. 
(2) The ejecta has a constant kinetic luminosity $\Lej(\tau)=L_0=10^{52}~\mbox{erg/s}$ for 
a duration of $\tau_b=5~\mbox{s}$, so that the total isotropic energy of the burst is 
$E_b=L_0\,\tau_b=5 \times 10^{52}~\mbox{erg}$. 
(3) The ejecta is assumed to emerge with a constant Lorentz factor $\Gej=300$, 
so that the RS is short-lived. 
(4) The burst is assumed to be located at a redshift $z=1$. 
As in U12 (Equation (35)), a flat $\Lambda$CDM universe is adopted for 
the luminosity distance, with the parameters $H_0=71$ km $\mbox{s}^{-1}$ $\mbox{Mpc}^{-1}$, 
$\Omega_{\rm m}=0.27$, and $\Omega_{\Lambda}=0.73$ (the concordance model). 
(5) The microphysics parameters are adopted as $p=2.3$, $\epsilon_e=10^{-1}$, 
and $\epsilon_B=10^{-3}$ (for slow cooling) or $10^{-1}$ (for fast cooling).
In the following, we perform calculations under the assumptions of two different
pressure profiles across the blast wave: a constant profile (\S2.1) and a
Blandford-McKee (BM) profile (\S2.2).

\subsection{Constant profile of pressure}

We find the blast wave dynamics of the fiducial example above, 
by making use of \cite{uhm11} and U12 (Section 2). 
It is found that the RS is short-lived, as expected, vanishing at around $\tobs=20$ s. 
Once the RS has crossed the ejecta, the blast wave deceleration transitions to and then 
agrees with the BM solution. During the transition phase, the deceleration (viewed as 
a function of radius $r$) is slightly slower than the BM solution,\footnote{This implies 
that, during the transition phase, the blast wave deceleration (now viewed as a function 
of observer time $\tobs$) is slightly faster than the BM case. For this reason, 
in Figure~\ref{fig:f3} below, the separation between two injection breaks at $10^2$ s 
and $10^3$ s is slightly wider than that expected from the BM deceleration.} 
since the pressure profile is being lowered into the BM profile and 
thermal (internal) energy contained there is being adiabatically converted into kinetic bulk motion. 
As the RS is short-lived here, we focus on the FS afterglow spectra from the FS shocked 
region (i.e., region 2). The magnetic field $B^i(t^{\prime})$ of $\delta m^i$ 
at any co-moving time $t^{\prime}$ ($t_i^{\prime}<t^{\prime}<t_j^{\prime}$) is found 
following U12 (Section 3.1) while assuming, for simplicity, a constant profile of pressure 
over region 2, i.e., $p^i(t^{\prime})=\pf(t^{\prime})$ for all shocked shells $\{\delta m^i\}$ 
in region 2 at every co-moving time $t^{\prime}$. Here, $\pf$ is the pressure at the FS front. 
Integrating Equation (\ref{eq:gamma_e}) from $t_i^{\prime}$ to $t_j^{\prime}$ with the initial 
values at $t_i^{\prime}$ (Equation (\ref{eq:initial})), we obtain the Lorentz factors 
$\gm^i (t_j^{\prime})$ and $\gc^i (t_j^{\prime})$ of each shell $\delta m^i$ at time 
$t_j^{\prime}$.

Taking also into account the equal-arrival-time ``curvature effect'' 
of spherical shells, the Doppler boosting of radial bulk motion, 
and different radii $\{r^i\}$ of individual shells (U12, Section 3.3), 
we calculate the FS afterglow spectra for the fiducial example above. 
The result is shown in Figure~\ref{fig:f1}. For panels (a) and (b), 
we use $\epsilon_B=10^{-3}$ (slow cooling), and for panels (c) and (d), we use 
$\epsilon_B=10^{-1}$ (fast cooling). The upper panels (a) and (c) show 
the instantaneous afterglow spectra at observer times $\tobs = 10^1, 10^2, 10^3, 10^4$, 
and $10^5$ s, and the lower panels (b) and (d) show a time evolution of spectral indices 
$\beta$ (with a convention, $F_{\nu} \propto \nu^{-\beta}$) at 1 keV and {\it R} band. 
As shown in panel (a), the afterglow spectra displays a very mild and smooth transition 
over a few orders of magnitude in frequency or energy in the vicinity of a cooling break. 
In particular, the spectral indices shown in panel (b) highlight that 
a transition from $\beta=(p-1)/2$ spectral regime to $\beta=p/2$ segment is very slow and 
takes several orders of magnitude in observer time $\tobs$. 
The outcome in panels (c) and (d) (fast cooling) is even more interesting. 
An expected transition from $\beta=-1/3$ to $\beta=1/2$ is not present. The characteristic 
spectral segment with $\beta=1/2$ is not even reproduced. Instead, an almost flat 
($\beta \approx 0$) spectral segment is observed at around the optical bands.

In an effort of identifying the main source of this smoothness, 
we first remove the curvature effect and place all the shocked shells at the same 
radius $\rf$ where the FS is located. 
All other steps of the calculations remain the same as above.
We find the afterglow spectra again for the fiducial example, and show the result 
in Figure~\ref{fig:f2}. As shown in panel (a), the injection break $\nu_m$ is now 
sharper than in Figure~\ref{fig:f1}, which can also be seen in panel (b) by 
noticing that the transition from $\beta=-1/3$ to $\beta=(p-1)/2$ regime is 
faster than in Figure~\ref{fig:f1}. The injection break $\nu_m$ 
in panels (c) and (d) (fast cooling) is also more visible than in Figure~\ref{fig:f1}; 
panel (d) exhibits a sharper feature between $\beta=1/2$ regime and $\beta=p/2$ regime. 
However, the afterglow spectra in panel (a) still exhibit a very mild and smooth 
cooling break. The spectral indices in panel (b) show that the transition 
from $\beta=(p-1)/2$ to $\beta=p/2$ segment is just as slow as in Figure~\ref{fig:f1} 
and still takes several orders of magnitude in observer time $\tobs$. 
Thus, this is also an indication that the curvature effect is nearly negligible in 
smoothing out a break if the break is already sufficiently smooth.

For panels (c) and (d) (fast cooling), the expected transition 
from $\beta=-1/3$ to $\beta=1/2$ regime is not present again; 
the characteristic fast-cooling segment with $\beta=1/2$ is not reproduced. 
The nearly flat ($\beta \approx 0$) spectral segment at around the optical bands 
is now more visible than in Figure~\ref{fig:f1}. 
This is because the standard fast-cooling synchrotron spectrum with index $\beta=1/2$ 
below the injection frequency \citep{sari98}, which has been widely believed, 
is not valid for a general problem involving a non-steady state electron spectrum of 
the emitting region, which we recently showed in the context of GRB prompt emission \citep{uhm13}.
Briefly speaking, the standard fast cooling spectrum is valid only for a steady state 
electron distribution, which may be achieved when a constant strength of magnetic field 
is invoked for the emitting region. 
However, in a rapidly expanding source such as for GRB prompt emission, 
the magnetic field strength in the emitting region cannot be preserved as a constant. 
The same is also true for GRB afterglow radiation, since the blast wave is decelerating. 
\cite{uhm13} showed that, for such a system with a decreasing magnetic field 
in the region, the fast cooling electron distribution is not in a steady state, 
forming a significantly harder spectrum than the standard one. The resulting 
synchrotron photon spectrum is also significantly harder than the standard one with 
$\beta=1/2$, naturally yielding a nearly flat ($\beta \approx 0$) spectral segment below 
the injection frequency. A more detailed investigation regarding this fast cooling spectrum 
in the context of GRB afterglow emission will be presented in a different paper.

Since the curvature effect is not the main source of smoothing the cooling break,
the next source we suspect is the adiabatic 
cooling term included in Equation (\ref{eq:gamma_e}). Dropping out this term from 
Equation (\ref{eq:gamma_e}) and integrating it from $t_i^{\prime}$ to $t_j^{\prime}$, 
we get 
\be
\label{eq:no_adiabatic}
\frac{1}{\gc^i (t_j^{\prime})} = \frac{\sigma_T}{6\pi m_e c} 
\int_{t_i^{\prime}}^{t_j^{\prime}} \left[B^i(t^{\prime})\right]^2 dt^{\prime},
\ee
for $\gc^i (t_j^{\prime})$ at time $t_j^{\prime}$, since $\gc^i (t_i^{\prime})=+\infty$.
For the same fiducial example, we calculate the afterglow spectra while making use 
of Equation (\ref{eq:no_adiabatic}) instead of Equation (\ref{eq:gamma_e}) in the 
calculation of $\gc^i (t_j^{\prime})$. The resulting afterglow spectra are not very 
different from those shown in Figure~\ref{fig:f2}. Thus, the adiabatic cooling term 
in Equation (\ref{eq:gamma_e}) is not the main source of this slow transition through 
the cooling break.

The analytical method of \cite{sari98} uses the instantaneous $B$ to estimate the
cooling time scale, which implicitly assumes that the magnetic field strength $B$
did not evolve during the dynamical evolution of the blast wave. To check whether
this is the main source of discrepancy, we take the $B^2$ term out of the integration 
in Equation (\ref{eq:no_adiabatic}). Using its current value $B^i(t_j^{\prime})$ 
at time $t_j^{\prime}$, we get 
\be
\label{eq:spn}
\frac{1}{\gc^i (t_j^{\prime})} = \frac{\sigma_T}{6\pi m_e c} 
\left[B^i(t_j^{\prime})\right]^2 (t_j^{\prime}-t_i^{\prime}),
\ee
which resembles the widely used expression 
for $\gc$ of Sari et al. (1998), their Equation (6). 
Employing Equation (\ref{eq:spn}) in the calculation of $\gc^i (t_j^{\prime})$ 
and, for simplicity, adopting $\gm^i (t_j^{\prime})=\ginj (t_j^{\prime})$ 
(Equation (3)) in all shocked shells $\{\delta m^i\}$, 
we find the FS afterglow spectra for the fiducial example. 
The result is shown in Figure~\ref{fig:f3}. Panel (a) now displays a considerably 
sharp cooling break. Thus, using Equation (\ref{eq:spn}) and based on the Lagrangian 
description of the blast wave, we closely reproduce the analytical afterglow spectra 
shown in Sari et al. (1998). On the other hand, panels (c) and (d) (fast cooling) 
indicate that the predicted spectral segment with $\beta=1/2$ is still not recovered. 
\cite{uhm13}, in particular, our Model [a] there, reproduced the standard 
fast cooling spectrum with $\beta=1/2$, (i) by assuming a constant strength of 
magnetic field $B$ in the emitting region, (ii) by adopting a constant injection 
rate $R_{\rm inj}$ of accelerated electrons, and (iii) by essentially following cooling 
of all the electrons inside the spectrum individually. Here (i) and (ii) are to assure 
that the global electron spectrum of the emitting region remains in a steady-state, and 
(iii) is to accurately follow a time evolution of the entire electron spectrum shape in 
each injection shell. Equation (\ref{eq:spn}) would play a similar role as (i) does, 
since $B$ is taken out of the time integration. However, for this afterglow calculation, 
we do not keep a constant injection rate $R_{\rm inj}$, since it is determined 
by the blast wave dynamics. Also, as mentioned earlier, we do not follow cooling of 
all the electrons inside the spectrum. For simplicity, we only follow the time evolution 
of two Lorentz factors $\gm^i$ and $\gc^i$ for each shell $\delta m^i$, and then assume 
that the shell $\delta m^i$ maintains a power-law distribution of slope $p$ between 
$\gm^i$ and $\gc^i$. These two issues (ii) and (iii) would become potential reasons 
why the standard spectral segment $\beta=1/2$ is not fully recovered here. 
This will be further investigated in a different paper.

Thus, we have identified the main source of the smooth cooling break. 
The approach made above to get Equation (\ref{eq:spn}) from 
Equation (\ref{eq:no_adiabatic}) cannot be justified, since the magnetic field strength 
$B^i$ in the shell $\delta m^i$ must be a time-dependent quantity. This requires 
an integration of $B^{i\,2}$ over time in order to correctly follow the cooling 
history of electrons in the shell $\delta m^i$, 
as is described by Equation (\ref{eq:no_adiabatic}).

\subsection{BM profile of pressure and Lorentz factor}

After the RS crosses the ejecta, it is expected that the blast wave
would be adjusted to the Blandford-McKee (1976) self-similar profile
\citep{kobayashi00}. 
\cite{granotsari02} also obtained smooth spectral breaks in their detailed
numerical modeling. Their model invokes the evolution of magnetic fields 
in the emission region, curvature effect, and the BM density/Lorentz factor 
profile of the blast wave. In order to investigate the effect of BM profiles 
on the shape of afterglow spectra, we now take into account the BM profiles 
in our numerical calculations. The fiducial example above has a short-lived RS. 
Thus, once the RS crosses the end of the ejecta, we fully adopt the BM profiles 
in our numerical code.

For a power-law profile of the ambient medium density, $\rho_1(r)/m_p=n_1(r) \propto r^{-k}$,
when the FS front is located at radius $r_f$ with the Lorentz factor $\Gamma_f$, 
a shocked fluid element at radius $r^i$ ($\leq r_f$) is described 
by the BM solution \citep{blandford76} as follows,
\beq
\label{eq:BM_pressure}
p      &=& \frac{2}{3}\, \rho_1(r_f)\, c^2\, \Gamma_f^2\, \chi^{-\frac{17-4k}{3(4-k)}}, \\
\label{eq:BM_Lorentz}
\gamma &=& \frac{1}{\sqrt{2}}\, \Gamma_f\, \chi^{-\frac{1}{2}}, \\
\label{eq:BM_num_den}
n      &=& 2\sqrt{2}\, n_1(r_f)\, \Gamma_f\, \chi^{-\frac{10-3k}{2(4-k)}},
\eeq
where the coordinate $\chi$ of the fluid element is given by 
\be
\label{eq:BM_chi}
\chi = 1+2(4-k)\, \Gamma_f^2\, \left( 1-r^i/r_f \right).
\ee
The fluid element has pressure $p$ and number density $n$ in the co-moving frame 
and moves with the Lorentz factor $\gamma$ in the lab frame.
For an adiabatic blast wave, this fluid element flows adiabatically and 
should satisfy $p \propto n^{4/3}$. Also, the FS front satisfies 
$\Gamma_f^2 \propto r_f^{k-3}$ if the blast wave is adiabatic. 
Then, together with Equations (\ref{eq:BM_pressure}) and (\ref{eq:BM_num_den}), 
the relation $p \propto n^{4/3}$ results in a simple and useful 
relationship $\chi \propto r_f^{4-k}$ (see also \cite{granotsari02}). 
In connection with our Lagrangian description of the blast wave, 
when a shell $\delta m^i$ is created at the FS front at lab time $t_i$, 
the $\chi^i$ coordinate of $\delta m^i$ at time $t_i$ is $\chi^i(t_i)=1$ 
since its radius is equal to the FS radius $r_f(t_i) \equiv r_f^i$. 
Then, the $\chi^i$ coordinate of the shell $\delta m^i$ 
at a later lab time $t$ ($\geq t_i$) is given by 
\be
\label{eq:BM_chi2}
\chi^i(t) = \left[ \frac{r_f(t)}{r_f^i} \right]^{4-k}.
\ee
Using Equations (\ref{eq:BM_chi}) and (\ref{eq:BM_chi2}), we then find 
the radius $r^i$ of $\delta m^i$ at time $t$ ($\geq t_i$) as 
\be
\label{eq:BM_radius}
r^i(t) = r_f(t) \left[ 1-\frac{\chi^i(t)-1}{2(4-k)\, \Gamma_f^2(t)} \right],
\ee
where $\Gamma_f(t)$ is the FS Lorentz factor at time $t$. 
Also, Equations (\ref{eq:BM_pressure}) and (\ref{eq:BM_Lorentz}) give 
the pressure $p^i(t)$ and the Lorentz factor $\gamma^i(t)$ of the shell $\delta m^i$ 
at time $t$ ($\geq t_i$), respectively,
\beq
\label{eq:BM_pressure2}
p^i(t)      &=& p_f(t)\, \left[\chi^i(t)\right]^{-\frac{17-4k}{3(4-k)}}, \\
\label{eq:BM_Lorentz2}
\gamma^i(t) &=& \frac{1}{\sqrt{2}}\, \Gamma_f(t)\, \left[\chi^i(t)\right]^{-\frac{1}{2}}.
\eeq

\begin{figure}
\begin{center}
\includegraphics[width=17cm]{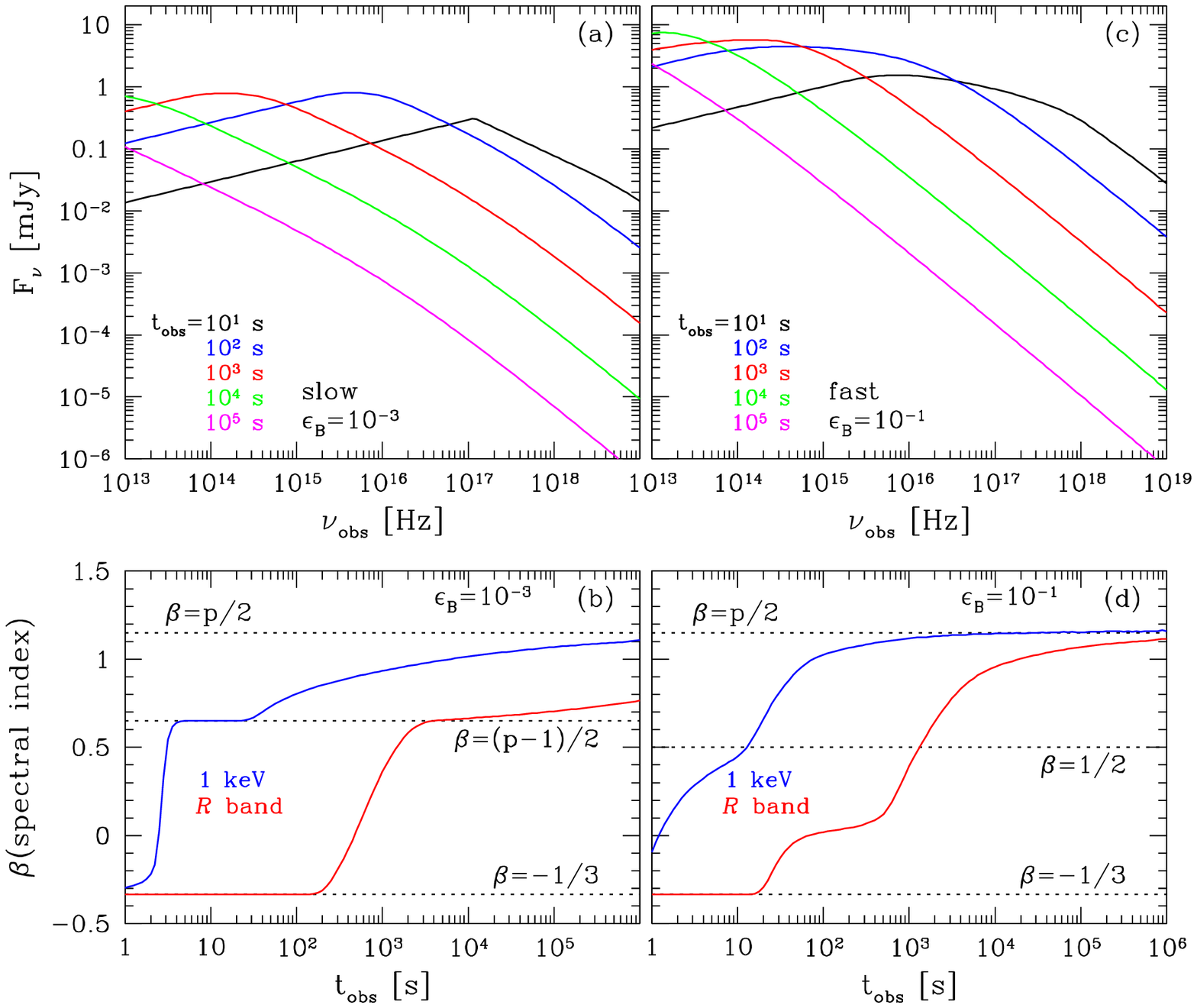}
\caption{
Afterglow spectra of the fiducial example at different 
observation times. A constant profile of pressure is adopted 
for the shocked blast region. 
Equation (1) is used to calculate the cooling
history of electron energies. The curvature effect is also fully 
taken into account. Following parameters are adopted: $\Lej(\tau)
=L_0 = 10^{52}~{\rm erg/s}$, $\tau_b = 5$ s,
$E_b = 5\times 10^{52}~{\rm erg}$, $n_1=1~{\rm cm^{-3}}$, 
$\Gej = 300$, $z=1$, $p=2.3$, $\epsilon_e = 0.1$.
{\em Top Left:} slow cooling case with $\epsilon_B = 0.001$.
{\em Bottom Left:} spectral index $\beta$ evolution as a function
of observer time $\tobs$ for the slow cooling case, 
which shows a very slow transition of $\nu_c$ crossing.
{\em Top Right:} fast cooling case with $\epsilon_B = 0.1$.
{\em Bottom Right:} $\beta$ evolution for the fast cooling case.
} 
\label{fig:f1}
\end{center}      
\end{figure}

\begin{figure}
\begin{center}
\includegraphics[width=17cm]{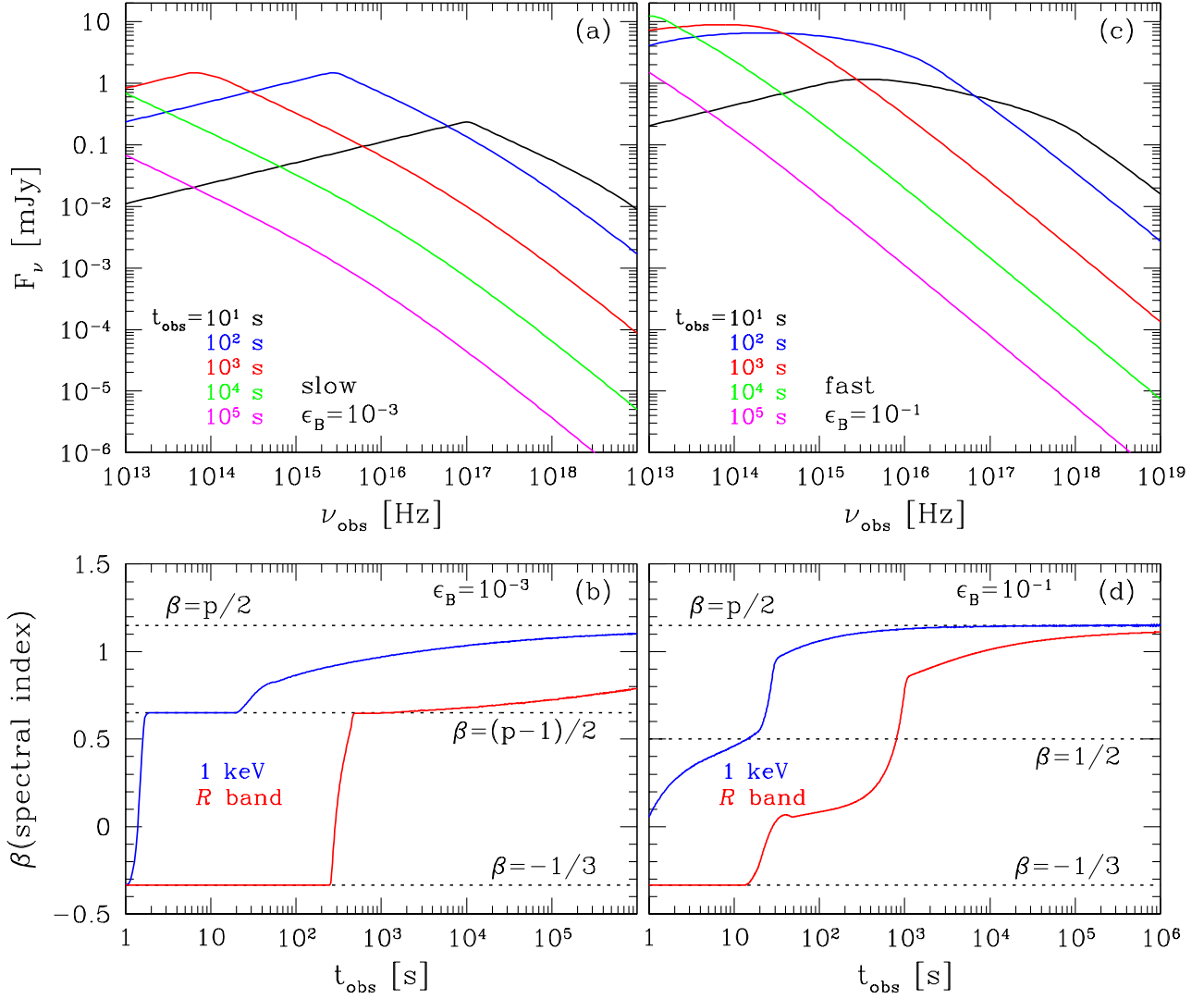}
\caption{
Calculation of the same problem as Figure~\ref{fig:f1}, 
but without including the curvature effect to allow 
identification of the key physical origin of $\nu_c$ smoothing. 
The notations of all 4 panels are the same as Figure~\ref{fig:f1}. 
} 
\label{fig:f2}
\end{center}      
\end{figure}

The fiducial example has $k=0$ (a constant density medium). Adopting the BM profiles 
described above, we calculate the afterglow spectra again for the fiducial example.
First, we find the magnetic field $B^i(t)$ of the shell $\delta m^i$ at time $t$ 
($\geq t_i$) following U12 (Section 3.1) but using the BM pressure profile 
(Equation (\ref{eq:BM_pressure2})) rather than the constant profile of pressure. 
Second, as we did in Figure~\ref{fig:f1}, we fully include the curvature effect 
(spherical curvature of shells, the Doppler boosting of radial motion, and different 
radii of shells). Equation (\ref{eq:BM_radius}) gives the radii of shells for the BM profile. 
Also, due to the BM Lorentz factor profile (Equation (\ref{eq:BM_Lorentz2})), each shell has 
its own Doppler boosting factor. Thus, we accordingly modify the synchrotron photon frequency 
and spectral flux density that are emitted from the shell. 
Third, we use Equation (\ref{eq:gamma_e}) to correctly follow cooling of electrons. 
The co-moving clock $t^{\prime}$ in a shell flows now differently from clocks in other shells. 
Taking this effect into account, we integrate Equation (\ref{eq:gamma_e}) from $t_i$ to $t_j$ 
($> t_i$) with the initial values at $t_i$ (Equation (\ref{eq:initial})) and obtain 
the Lorentz factors $\gm^i$ and $\gc^i$ for the shell $\delta m^i$ at time $t_j$.

\begin{figure}
\begin{center}
\includegraphics[width=17cm]{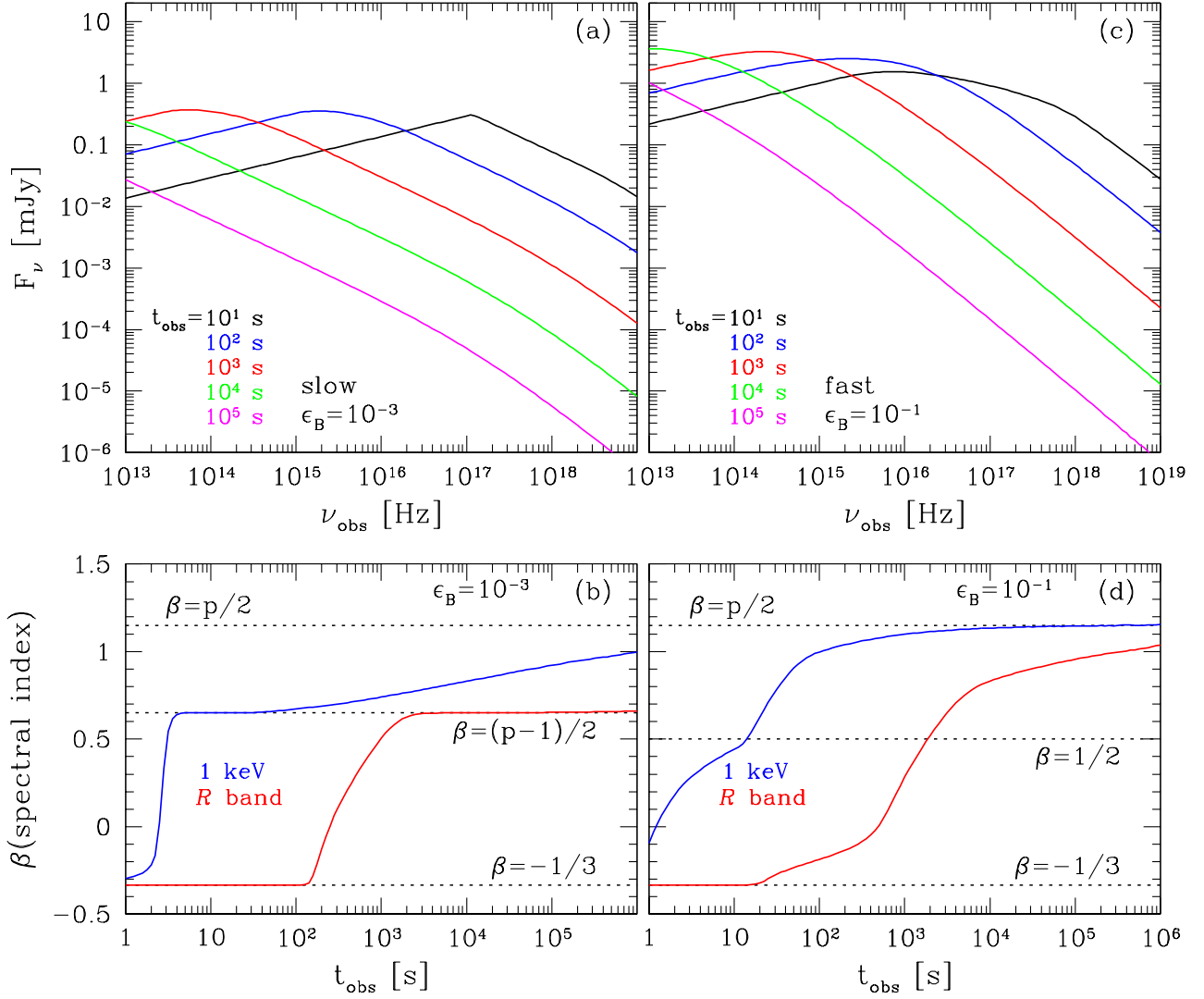}
\caption{
Calculation of the same problem as Figure~\ref{fig:f1}, but with 
the BM profiles for the blast wave. The notations of all 4 panels 
are the same as Figure~\ref{fig:f1}. 
} 
\label{fig:f4}
\end{center}      
\end{figure}

\begin{figure}
\begin{center}
\includegraphics[width=17cm]{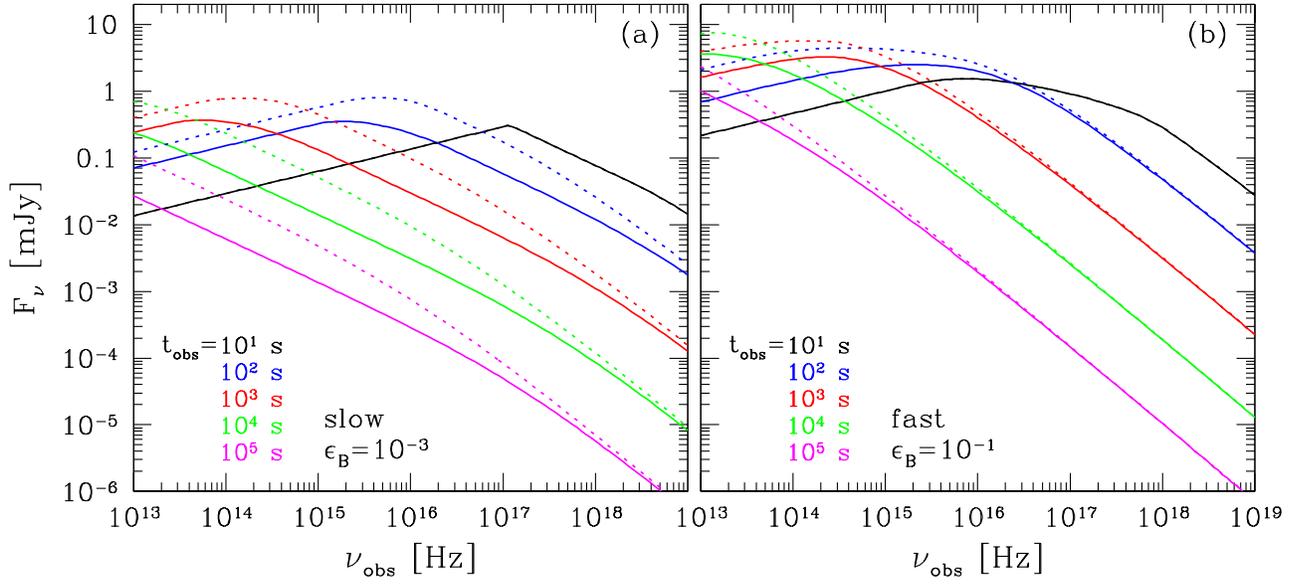}
\caption{
Comparison between two different profiles: 
solid line for spectra from Figure~\ref{fig:f4} (BM profile) 
and dotted line for spectra from Figure~\ref{fig:f1} (constant profile). 
{\em Left:} $\epsilon_B=0.001$ (slow cooling). 
{\em Right:} $\epsilon_B=0.1$ (fast cooling). 
} 
\label{fig:f5}
\end{center}      
\end{figure}

The resulting afterglow spectra is shown in Figure~\ref{fig:f4}. 
As shown in panel (a), the spectra still exhibit a very mild and smooth cooling break $\nu_c$. 
The spectral indices in panel (b) show that the transition from $\beta=(p-1)/2$ regime 
to $\beta=p/2$ segment is still as slow as in Figure~\ref{fig:f1} (constant profile of pressure), 
but occurs later than in Figure~\ref{fig:f1} indicating that the cooling break is located 
at higher energy than in Figure~\ref{fig:f1}. The outcome in panels (c) and (d) (fast cooling) 
is similar to that of Figure~\ref{fig:f1}, except that the flat ($\beta \approx 0$) spectral 
segment at around the optical bands is less prominent than in Figure~\ref{fig:f1}. 
For comparison, we show these two results together in Figure~\ref{fig:f5}: solid line 
for spectra from Figure~\ref{fig:f4} (BM profile) and dotted line for spectra 
from Figure~\ref{fig:f1} (constant profile). 
The left panel is for $\epsilon_B=10^{-3}$ (slow cooling), and the right panel is 
for $\epsilon_B=10^{-1}$ (fast cooling). 
This fiducial example has a short-lived RS, which vanishes at around $\tobs=20$ s. 
Therefore, two results are identical to each other at $\tobs=10$ s. For the spectra 
after the RS is vanished, the general trend is that the dotted lines have higher 
spectral flux $F_{\nu}$ than the solid lines, because the blast wave with 
a constant profile of pressure has higher energy density and magnetic field 
than the blast wave with a BM profile of pressure. However, the difference 
in spectral flux $F_{\nu}$ becomes less significant as we go to higher photon 
frequency $\nuobs$. This is because only recently shocked fresh shells have 
electrons with higher energy, which could contribute to higher frequency range 
of spectra. As for these fresher shells, the difference between two profiles 
becomes less significant. In particular, for the right panel (fast cooling), 
two results are essentially identical to each other at high frequency, 
indicating that only very fresh shells contribute to this energy range. 
For the left panel (slow cooling), the dotted lines have an injection break 
at higher frequency than the solid lines, because the dotted lines with 
constant pressure profile underestimate the degree of adiabatic cooling of electrons 
when compared to the solid lines with the BM profile; see the adiabatic cooling 
term in Equation (\ref{eq:gamma_e}). 
Also, from the left panel (slow cooling), one may notice that the solid lines 
have a cooling break at higher frequency than the dotted lines, as mentioned above. 
This is because the dotted lines have higher magnetic field (namely, higher rate of radiative cooling 
of electrons) than the solid lines. 
Nevertheless, the smoothness of cooling break in two results is comparable 
to each other (also shown in panel (b) of Figure~\ref{fig:f4}). 
This again suggests that the magnetic field evolution effect is the main
physical origin for the smooth cooling breaks.

%
%

\section{Conclusions and Discussion} \label{section:discussion}

In this paper, we perform a detailed study on the formation of GRB 
afterglow spectra, by adopting a Lagrangian description of the shocked region 
following \cite{uhm12}.
We precisely follow the cooling history of the electron spectrum for each individual 
Lagrangian shell, and integrate over all shells to find the instantaneous flux spectra.
We show that this detailed calculation gives rise to a very mild and smooth cooling break, 
which occurs gradually over a few orders of magnitude in energy. 
We identify the main source of this slow transition as due to the different
$B$ evolution histories of different mini shells. This gives rise to an
additional spreading of $\nu_c$ for different shells, aside from the
simple age difference ($t_j^{\prime} - t_i^{\prime}$). This extra spreading 
is the main source of the smooth $\nu_c$.

We have shown that this effect exists regardless of whether the curvature effect 
of a relativistic spherical shell is taken into account. 
It does not depend on the details of the blast wave dynamics or whether it is FS or RS.
In fact, it is an intrinsic effect relevant to a wide range of astrophysical phenomena 
that invokes synchrotron cooling of electrons.


It is interesting to note that the injection break $\nu_m$ is always much sharper than 
$\nu_c$ for slow cooling, and is usually so for fast cooling, even though it becomes 
somewhat smoother when the curvature effect is taken into account. 
If a sharp spectral break is observed in an astrophysical phenomenon, 
this break is very likely an injection break, and cannot be a cooling break.

Simple analytical GRB afterglow models predict some ``closure relations'' between
the temporal decay index $\alpha$ and the spectral index $\beta$
\citep[e.g.][]{meszarosrees97,sari98,sari99,chevalier00,daicheng01,zhangmeszaros04,gao13}.
These models have been applied to the Swift XRT data (from which both temporal
decay index and spectral index can be extracted) to test the validity of the
afterglow models \citep[e.g.][]{liang07b,liang08,panaitescu07,willingale07,evans09}.
The result of this paper suggests that the analytical closure relation above $\nu_c$ is 
usually not achieved. The data could still be consistent with the afterglow theory
even if the data fall into the grey zone between the analytical 
$\nu<\nu_c$ and $\nu>\nu_c$ closure relation lines in the $\alpha-\beta$ plane.

%
%

\acknowledgments 

We thank the anonymous referee for valuable comments and suggestions,
which allowed us to significantly improve the presentation of the paper.
We acknowledge Kavli Institute for Astronomy and Astrophysics, Peking University, 
for hospitality where this research is carried out.
BZ acknowledges a Cheung Kong Scholarship in China. 
This work is partially supported by NSF AST-0908362.

%
%


%
%

\begin{figure}
\begin{center}
\includegraphics[width=17cm]{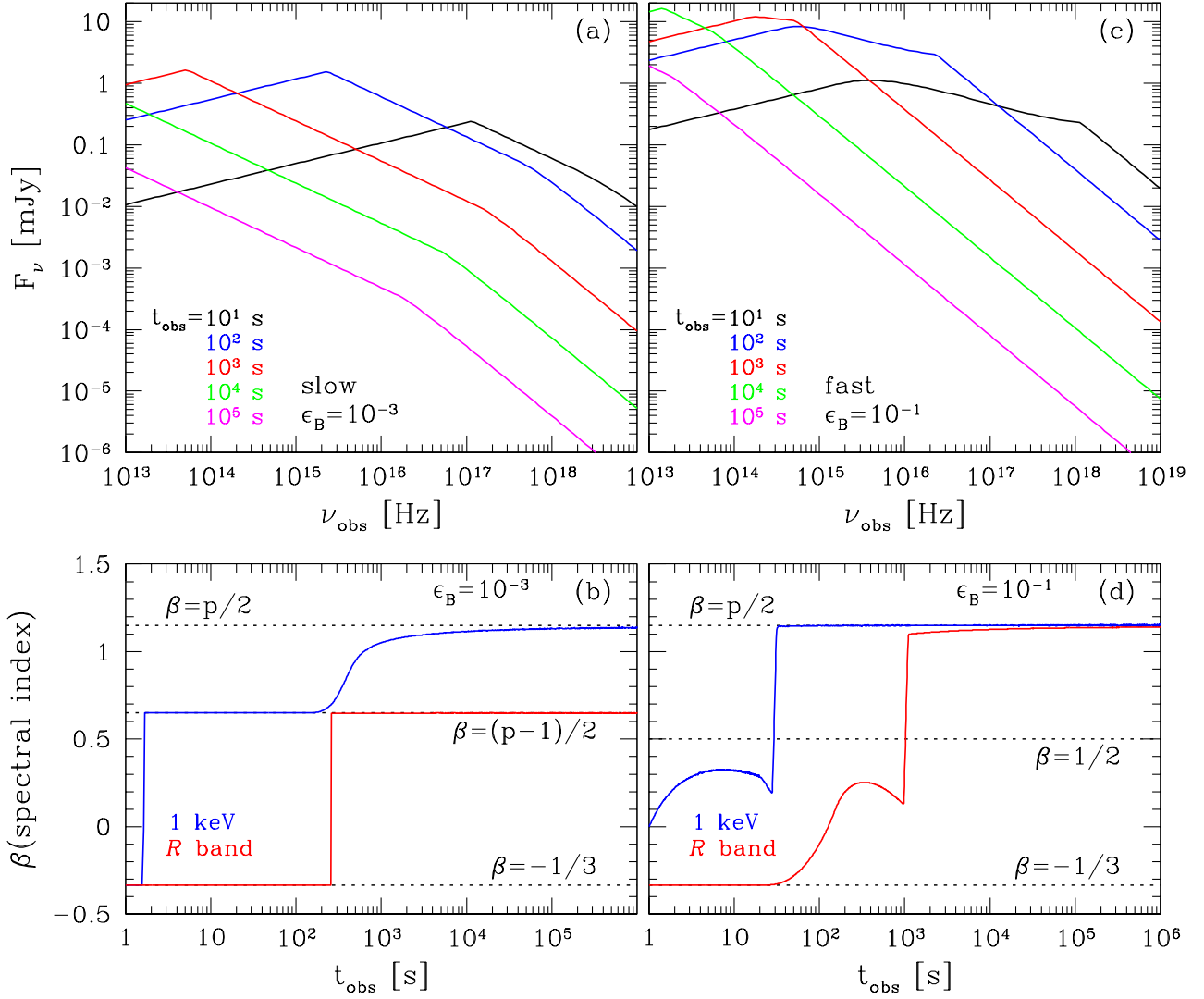}
\caption{
Calculation of the same problem as Figure~\ref{fig:f2}, but with the simple, 
unjustified prescription as described in Equation (\ref{eq:spn}). 
The notations of all 4 panels are the same as Figure~\ref{fig:f1}. 
A sharp $\nu_c$ is reproduced.
} 
\label{fig:f3}
\end{center}      
\end{figure}

\end{document}